# Metastable Ultracondensed Solid Hydrogenous Materials

W. J. Nellis

*Harvard University, Department of Physics, Cambridge MA 02138, USA*

The primary purpose of this paper is to stimulate theoretical predictions of how to retain metastably hydrogenous materials made at high pressure $P$ on release to ambient. Ultracondensed metallic hydrogen has been made at 140 GPa at finite temperatures $T$ in the fluid. The term "metallic" here means quantum mechanically degenerate. A single sample of ultracondensed hydrogen has been made at an estimated pressure of 495 GPa at 5.5 K. Whether that sample is solid or fluid remains to be demonstrated. Those results imply the long quest for metallic hydrogen is likely to be concluded in the relatively near future. Because the quest for metallic hydrogen has been a major driver of high pressure research for decades, a logical question is whether another research direction, comparable in scale to that quest, will arise in high pressure research in the future.

One possibility is retention of metastable solid metallic hydrogen and other hydrogenous materials on release of $P$ and $T$ to ambient. If hydrogenous materials could be retained metastably in the solid on release, those materials would be a new class of materials for scientific investigations and technological applications. This paper is a brief review of the synthesis of metallic hydrogen, potential technological applications of metastable solid metallic H and other hydrogenous materials at ambient, and published experimental and theoretical results as general background for what has been accomplished with metastable phases in the past, which suggests what might be accomplished in the future.

I. Introduction

Metallic fluid H (MFH) has been made under quasi-isentropic compression at pressures $P$ and temperatures $T$ achieved with a reverberating shock wave [1-4]. $P$ and $T$ achieved in those dynamic experiments are in the ranges 93 to 180 GPa and 1700 to 3100 K, respectively [2]. That crossover completes at the density $\rho$ of the dissociative metallization transition at "very low temperatures" predicted by Wigner and Huntington (WH) in 1935 [5]. MFH has also been made at finite $T$ under static compression along the dissociative phase transition curve at $P$ and $T$ in the ranges 82 to 170 GPa and 2500 to 1100 K, respectively [6-8]. Pressures $P$ and $T$ achieved under those dynamic and static compressions are comparable, although the trajectories in $P$-$T$ space achieved under those static and dynamic compressions are different because they were derived using different experimental phenomena [4]. A single observation of metallic H has been reported under static compression at estimated 495 GPa and 5.5 K [9], a result that needs to be reproduced and the phase of the sample, fluid or solid, needs to be determined. Although that sample is thought to be a solid, Babaev et al have predicted metallic hydrogen is a fluid near T = 0 K [10].

One conclusion implied by those experiments is that ultracondensed metallic hydrogen has been made in the fluid and is likely soon to be made in the solid. Thus, it appears that the decades-long quest for metallic hydrogen, which was launched by WH in 1935 and initiated experimentally in 1956 [11], is essentially reaching conclusion in the relatively near future. While metallic H is not the only important direction in high pressure research, it has been an important driver of this field for several decades.

A logical question then is whether another challenging research direction,

comparable in scale to the search for metallic H, will arise. One such possibility is retention of metastable solid metallic hydrogen on release of extreme pressures. This project will likely require at least as much time to accomplish as finding metallic H and thus the time is appropriate to start thinking about how it might be accomplished.

Recent achievements of ultracondensed hydrogenous materials at extreme dynamic conditions with lifetimes ranging from 10 to 100 ns have opened up newly accessible areas of research in condensed-matter physics, chemistry, materials science and planetary interiors [**4**]. If such hydrogenous materials made at high pressures, densities and temperatures could be retained as dense metastable solids on release to ambient, then those materials would be a new class of materials for scientific and technological investigations. For example, MFH with H density 9-times greater than that of H in liquid $H_2$ at 20 K has been made under dynamic compression. If MFH could be quenched to a metastable amorphous solid at ambient, then traditional condensed-matter experiments could be performed to measure structural, electrical, thermal, magnetic, superconducting and other properties of those recovered solids.

This opportunity depends on quenching hydrogenous fluids from high $P$, $\rho$ and $T$ to ambient, which is a significant challenge. *The primary purpose of this paper is to stimulate theoretical predictions of how to retain metastably hydrogenous materials made at extreme P and retained on release to ambient.* To demonstrate that the time is right to learn how to make metastable materials, a few experimental results for hydrogenous materials under dynamic and static compression are listed below and potential scientific investigations and technical applications of metastable hydrogenous materials are given. The use of intuition to choose successful experiments to determine systematics of achieving metastable hydrogenous materials is not expected by this author to be particularly fruitful in the foreseeable future. Theoretical predictions are needed to guide experiments.

Metastable dense hydrogenous materials have the potential to affect life as we know it in terms of (i) high-$T_c$ superconductors for efficient electrical energy transmission; (ii) quantum solids with unusual physical properties at room temperature, (iii) clean fuels for autos and other vehicles, (iv) energetic propellants for rocket-driven space travel, (v) energy-storage media, (vi) light-weight structural materials, and (vii) nuclear fuel in the isotopic forms of deuterium (D) and tritium (T) for Inertial Confinement Fusion (ICF), a potential source of commercial energy [**12**].

II. Some Hydrogen Results at High Pressures

In 1935 Wigner and Huntington (WH) predicted that ultracondensed $H_2$ would dissociate to H at a density $\rho = 0.62$ mol $H/cm^3$, $P$ greater than 25 GPa and "very low temperatures" $T$ [**5**]. Dense diatomic $H_2$ is an insulator with two electrons localized on each $H_2$ molecule. Dense monatomic H has a half-filled electronic energy band and thus is a metal. WH's prediction initiated a multi-decade search for metallic and superconducting hydrogen and its alloys.

Pressures $P$ up to several 100 GPa (100 GPa = $10^6$ bar = 1 Mbar), compressions $\rho/\rho_0$ up to ~10 fold, where $\rho$ is compressed density and $\rho_0$ is initial density, and temperatures $T$ up to several 1000 K in H can be accessed experimentally. Temperatures $T$ of a metal are low or high relative to its Fermi temperature $T_F$ [**13**]. Free-electron $T_F \propto \rho^{2/3}$. Because hydrogen is very compressible, low degeneracy factors $T/T_F$ are readily obtained. For metallic H with $\rho/\rho_0 = 10$, $T_F \propto 4.7 \rho_0$. As a result, ultracondensed matter ($T/T_F \ll 1$) and warm dense matter ($T/T_F \sim 1$) of hydrogen and hydrogenous materials are readily obtained with apparatus ranging in size from that of a two-stage light-gas gun (2SG ~10 m long) to a diamond anvil cell (DAC - hand-held). Substantially more extreme dynamic $P$, $\rho$, and $T$ are obtained with giant pulsed lasers and magnetically accelerated impactors.

If $H_2$ is completely dissociated and ionized at WH's predicted metallization density, the corresponding free-electron $T_F = 220,000$ K and so $T/T_F \approx 0.01$ at 2200 K. Thus, at WH's predicted metallization density, $T$s of a few 1000 K are low relative to $T_F$ and so finite $T$s are used to make dense metallic hydrogen. In the 1930s it was not possible to achieve the high pressures estimated to achieve 0.62 mol H/cm$^3$. So, WH simply calculated a lower bound, 25 GPa, of dissociation pressure by assuming that the compressibility of $H_2$ is pressure-independent [5]. Because compressibility generally decreases with pressure, measured metallization pressures of dense H are much greater than 25 GPa.

Metallic fluid H has been made with dynamic compression at 140 GPa, 0.63 mol H/cm$^3$, 3000 K and $T/T_F \approx 0.014$ [1,2] with measured Mott's Minimum Metallic Conductivity (~2000/(Ω-cm) [14]. The maximum measured melting temperature of $H_2$ in a DAC is 1060 K at 65 GPa, which decreases at lower and higher pressures [15]. At 3000 K and 140 GPa, hydrogen is a fluid. Those dynamic compressions were achieved with a two-stage light-gas gun (2SG) [16], which generated pressures in fluid H by impact of a plate at velocities as large as 7 km/s onto a cryogenic sample holder [2,17].

An observation of metallic solid H has been reported under static compression in a diamond anvil cell at 495 GPa and 5.5 K with a density substantially higher than 0.63 mol H/cm$^3$ [9]. Metallization pressure of ultracondensed hydrogen is extremely sensitive to temperature: 140 GPa at 3000 K in the fluid and ~500 GPa at 5 K in the solid or liquid. This result suggests that tuning *P/T* at intermediate conditions might produce an optimal quench to make metastable solid metallic H.

Ashcroft predicted that hydrogen alloys with a dominant constituent of H would become metallic at a lower pressure than pure H [18]. H$_2$S has been found to become a metal with a record-high superconducting critical temperature $T_c$ = 203 K at a static pressure of ~90 GPa [19], which is substantially lower than metallization pressure of 495 GPa reported for pure H [9]. Those results illustrate the value of theoretical predictions to guide experiments in a newly accessible thermodynamic regime.

III. Metastability: general considerations

Results described above imply numerous scientific and technological applications exist, if metastable solid dense hydrogenous materials could be retained on release of high *P/T* to ambient. At present it is not known how, nor even whether, a given high-pressure hydrogenous phase could be made to be metastable on release of pressure. The difficulty is illustrated by the dissociation and ionization energies of H$_2$, which are 4.5 eV and ~15 eV [20], respectively. *A key goal is development of systematic understanding of achieving metastability on release of pressure* – what is an optimal process and how can it be controlled and tuned? Such a tool would be extremely advantageous for making a range of metastable compositions and structures.

Scalability from research-size to commercially-attractive sample sizes is an important consideration in making metastable hydrogenous materials. Research-size sample dimensions speculated upon herein are as small as μm in thickness. Dynamic pressures are typically generated by an impactor accelerated up to several km/s or by irradiation by a modest size laser. In principle, sample volumes can be scaled up substantially if the appropriate dynamic compression process can be determined in small-scale research experiments. For example, dynamic pressures in ceramic powders up to ~100 GPa have been achieved with high-explosive-driven systems and associated sample holders ~1 m in diameter and ~4 m high by E. I. du Pont de Nemours & Co. [**21**].

IV. Possibilities

It took decades to make metallic H experimentally. Retaining metallic H at high metallization density on release of metallization pressure to zero is expected to be of comparable difficulty. In this section various types of hydrogenous and other materials are surveyed with respect to possible retention of metastable phases on release of pressure.

Various potential scientific and technological applications do not all require that metastable hydrogenous materials have high H densities required of metals. Electrical insulators, for example, are acceptable as energy-storage media. Thus, one approach might be to learn the systematics of making metastable hydrogenous materials and use H densities actually achieved metastably in various materials to determine uses of specific materials. The successful pressure release process might be a combination of several methods discussed below.

IV.1  Electronic band structure and H metastability

The electronic band structure of metallic H at $P = 0$ has been calculated and prospects for metastability considered [**22**]. That classic theoretical study found that metallic hydrogen likely crystallizes into anisotropic, filamentary, triangular structures with nearly degenerate energies and two-dimensional periodicity. The lifetime of the likely metastable ground state is finite but its value remains an open question, which is a serious issue for technological applications. The lifetime of a metastable state can probably be increased by application of a pressure that is much lower than the dissociative metallization pressure, which might not be a very low pressure from a practical standpoint. Near-degeneracy of likely structures suggests the possibility that degenerate solid H is highly defected or amorphous. In fact, metallic hydrogen has been predicted to be a liquid near T = 0 K [**10**].

IV.2  Nitrogen

A polymeric phase of nitrogen has been found in which all atoms are connected by single covalent bonds at pressures above 110 GPa and temperatures above 2000 K obtained in a laser-heated diamond anvil cell. At 300 K polymeric nitrogen is metastable down to 42 GPa. Polymeric nitrogen is predicted to have an energy content more than five times greater than existing energetic materials [**23**].

IV.3  Evolutionary structure searches and molecular dynamics

Layered $SiS_2$ structures at high pressures have been found with *ab initio* calculations that combine evolutionary structure searches and molecular dynamics [**24**].

Those computational methods appear to this author to be the most likely at present to identify possible ways to retain metastable hydrogenous materials made at high pressures, densities and temperatures and retained on release of those variables to ambient. Superconducting $H_2S$ with $T_c$ = 203 K at 90 GPa [19] and compounds with similar compositions and structures are a class of materials to investigate for high-$T_c$s that are possibly metastable on pressure release.

The capability of manipulating phase stability calculationally might be useful, for example, for designing materials that absorb high densities of H for use in energy-storage media. In this case the goal would be to facilitate absorption of high densities of H into a material, after which it ideally would be difficult to desorb H in order to maintain stable energy storage. Unless of course, desorption of H is desired for use in a fuel cell, in which case relatively little energy would ideally be required for desorption. Tuning ease of adsorption and desorption of H is a challenging materials issue.

IV. Silicon

Si is one of the few elements known to readily form a metastable phase naturally on release from high pressures. At room temperature, Si II (β-Sn body-centered tetragonal) transforms to Si III (BC-8 body centered cubic) as pressure releases to zero. Si BC-8 is a metastable phase [25,26]. Such a situation potentially enables tuning the ground state structure and its associated properties, provided the metastable phase is sufficiently long-lived.

IV.4 Films

Metallic fluid H has been made by supersonic hydrodynamics in ~100 ns by reverberation of dynamic pressure up to 140 GPa and 2600 K [2]. Initial thickness of the liquid $H_2$ sample layer was 500 μm, which decreased to ~50 μm under dynamic compression. Sample diameter was 25 mm. High *T*s can be generated by supersonic compression, which is followed by fast sonic release from maximum pressure to zero [27].

During release of *P* and *T* of MFH strong attractive forces between H atom pairs quickly recombine H atoms into $H_2$ molecules, which are not metallic. Thus, it appears to be necessary to quench MFH to an amorphous or glassy solid within ~100 ns after completion of synthesis of MFH. In that brief time interval it is not possible to quench monatomic MFH to amorphous metallic H by mechanical means. The quench of MFH to a solid must be designed into the real-time synthesis process.

Fast thermal quench occurs in films by fast thermal transport out of a thin specimen. μm-thick films must be encapsulated in thermally-conducting strong materials that can withstand the rapid stress release without allowing too much associated damage. Nb films a μm thick embedded in Cu and strong steel have been successfully recovered intact from 100 GPa dynamic pressures [28].

A method to fabricate superconducting high-Tc powders compacted dynamically into Cu strips has been proposed. The method uses an explosive-driven travelling shock wave to dynamically compress the assembly. The method potentially might be used to fabricate superconducting cable [29].

V. Conclusion

Determination of a process to systematically make metastable hydrogenous materials by quenching phases made at high pressures to ambient needs to be based on an extensive experimental database and associated theory, including the measured phase diagram of hydrogen at extreme $P$, $\rho$ an $T$. Mechanisms of appropriate phase transitions and hopefully determination of electronic structures that systematically induce metastability need to be determined. At present this author is not aware of a dynamic process that would accomplish these goals for metallic H. Neither does this author have a clear idea on how to begin to learn that process. Theoretical predictions are needed to stimulate experiments to begin learning that process.